\documentclass[a4paper,aps,twocolumn,nofootinbib]{revtex4}
\RequirePackage[colorlinks,hyperindex]{hyperref}
\RequirePackage[english]{babel}
\RequirePackage[latin1]{inputenc}
\RequirePackage[T1]{fontenc}
\RequirePackage{mathrsfs}
\RequirePackage{amsmath}
\RequirePackage{amssymb}
\RequirePackage{amsbsy}
\RequirePackage{color}
\RequirePackage{bm}
\hypersetup{colorlinks=true,breaklinks=true,urlcolor=blue,linkcolor=red}
\begin{document}
\title{\bf{Polar analysis of the Dirac equation through dimensions}}
\author{Luca Fabbri}
\affiliation{DIME Sez. Metodi e Modelli Matematici, Universit\`{a} di Genova,
Via all'Opera Pia 15, 16145 Genova ITALY}
\date{\today}
\begin{abstract}
We consider the polar form of the spinor field equation in an $n$-dimensional space-time, studying the way in which the space-time dimension influences the number of the independent field equations and the number of the degrees of freedom of the spinor field written in the polar form: we will find that this polar form is the clearest tool to make manifest the fact that the degrees of freedom of the spinor field and the independent field equations match in dimensions $4$ and $8$ alone while in all the other instances there will be problems of general under-determination or over-determination.
\end{abstract}
\maketitle
\section{Introduction}
Although general differential geometry is definable in any number of dimensions, nevertheless there are objects that are rather sensitive to the number of dimensions of the space where they are defined: a first example can be seen in the theory of differential forms, in which we have the Levi-Civita completely-antisymmetric pseudo-tensor defined with as many indices as the space itself. Thus, a completely antisymmetric object like the torsion can not be defined in $2$ dimensions, but it is dual to a scalar in $3$ dimensions, or dual to an axial-vector in $4$ dimensions.

Another case of dimensional dependence is the curvature tensor: the Riemann tensor is a fundamental object in $4$ dimensions, but it is determined by the Ricci tensor in $3$ dimensions, by the Ricci scalar in $2$ dimensions, and it is trivial in $1$ dimension. All this is rather well known.

The dependence on the dimension is perhaps even more dramatic in the case of the spinor fields \cite{L, b, c, Gogberashvili:2005cp}: as known, in $2$ dimensions the Clifford algebra gives $1$ generator for the Lorentz group, in $3$ dimensions it gives $3$ generators, in $4$ dimensions $6$ generators. But things are not always so linear, because in both $2$ and $3$ dimensions spinor fields have $2$ components, and only by going to $4$ or $5$ dimensions they have $4$ components, and so on doubling their components every time we increase by two dimensions.

What is interesting, though presumably not surprising, is the fact that this irregular jumping through dimensions of the Clifford algebra, on the one hand, and of the spinor field, on the other hand, renders such a sensitivity to the dimension quite erratic: as is well known, in $4$ dimensions the spinor fields have $4$ complex components, and thus $8$ real components, but only $2$ degrees of freedom, whereas in $2$ dimensions they have $2$ complex components, thus $4$ real components, with $3$ degrees of freedom. Such results are common knowledge for spinor classification in various dimensions as discussed in \cite{s, Lopes:2018cvu, daRocha:2006vzt, Cavalcanti:2014wia}, where some very general treatments are done which also include the enlargements to arbitrary signatures and up to octonion-valued fields.

A general discussion about real, complex, quaternionic and octonionic structures, for an arbitrary dimension and signature, is in \cite{Carrion:2003ve, Toppan:2004xy, Okubo:1990bf, Okubo:1991qj}: in these papers it is discussed that fundamental spinor fields are real for $n\!=\!2$ and $n\!=\!3$, they are complex for $n\!=\!4,\ 8$ and quaternionic or octonionic in other dimensions. As we are going to see next, similar results can also be obtained by employing another type of analysis involving what is known as polar form of the spinor field, and which consists in writing the spinor field as a function of some module and some angle.

The reason for the fact that spinor fields have such erratic behaviour through different dimensions is that they are defined by their transformation properties under the complex Lorentz group, so that components representing redundant information can be removed by means of the complex Lorentz transformation: so in going from $2$ to $4$ dimensions we increase from $4$ to $8$ the real components of the spinor fields but in addition we increase from $1$ to $6$ the Lorentz transformations that can be used to remove those redundancies, consequently reducing the amount of degrees of freedom. Once again, this fact is well known, but differently from the results obtained in the mentioned literature, the spinor erraticity can be easily understood using the previously introduced polar form, as it has been discussed in \cite{daSilva:2012wp, Bonora:2014dfa, Fabbri:2017lvu, Fabbri:2016msm}, in which we find the general study of the classification of degrees of freedom together with extensions to arbitrary dimensions and non-abelian cases.

But what is more interesting, and now presumably also surprising, is that such an erratic behaviour through dimensions of degrees of freedom of the spinor fields and of independent field equations do not correspond: they are both erratic, but in different ways, so that the degrees of freedom of the spinor fields and the independent field equations do not match. On the other hand, they should match since the degrees of freedom of a spinor field must be equal to the physical components determined by the independent spinorial differential field equations.

Consequently, the analysis of the dimensions in which spinor fields are defined can also be done in terms of the corresponding spinor field equations: clearly, all results anticipated above will be recovered. But intriguingly, all recovered results will be easily interpreted when the polar form of the spinor field equations is taken \cite{Rodrigues:1996tj, h1}.

The reason is simple: spinors that are well-defined have a number of degrees of freedom matching the number of field equations, so that the polar form will resolve into a perfect set of equations \cite{k, Fabbri:2016laz}, but spinors with problems in their definition will have some degrees of freedom in excess or in defect with respect to the number of field equations, and the polar form will manifest this situation by developing too few or too many equations. Therefore, the classification of degrees of freedom through dimensions that has been discussed in terms of Clifford algebras and spinor fields can additionally be achieved in terms of the Dirac spinor field equation, and then in polar form.

The advantage of converting the results from algebraic perspectives to differential perspectives is to give them a less kinematic and more dynamic interpretation, and the advantage of converting them into polar form is that all quantities will be written in terms of real tensors and so easier to be read: the Dirac equation in polar form is in fact perfectly suited to be decoded in terms of quantities that have a simple and clear meaning \cite{Fabbri:2018crr}. In polar form, problems like the abundance of physical components will be translated into the lack of the corresponding equations governing the dynamics of such extra components while problems like the lack of physical components is going to be converted into the fact that some field equations will collapse into constraints between quantities that should otherwise be freely defined, as we will see in this paper.
\section{The General analysis}
In the general $n$-dimensional space-times, the spinorial fields are defined as columns of 
$2^{(n-a)/2}$ complex scalars with $a$ being zero or one for $n$ even or odd; because these components are complex, we have $2^{(n-a+2)/2}$ real scalars as total number of components. In the definition of spinor fields as column of complex scalars we also have to add that they transform according to the complex representation of the Lorentz group, being $n(n\!-\!1)/2$ transformations in total. These transformations can be employed to remove the redundant components leaving only the true components, or degrees of freedom, in the spinor fields.

After this operation is done, the spinorial field remains with $2^{(n-a+2)/2}\!-n(n-\!1)\!/2$ components that can no longer be reduced, and in this form the spinor field is said to be in polar form: because these are the irreducible degrees of freedom, that is the physical fields, all of their derivatives must be specified by the field equations, and they amount to the number of  $[2^{(n-a+2)/2}\!-\!n(n\!-\!1)/2]n$ in total.

These conditions must come from the spinor field equations, which amount to a number of 
$2^{(n-a+2)/2}$ independent field equations in total. Because the number of independent field equations must not exceed the number of degrees of freedom of the spinor fields, we have that 
\begin{eqnarray}
&[2^{(n-a+2)/2}\!-\!n(n\!-\!1)/2]n\!\geqslant\!2^{(n-a+2)/2}
\end{eqnarray}
with the optimal case given by the strict equality.

Since the number of dimensions is strictly larger than one, this condition can be worked out to be
\begin{eqnarray}
&2^{(n-a+4)/2}\!-\!n^{2}\!\geqslant\!0
\label{cond}
\end{eqnarray}
again with the optimal case given by the strict equality, and which we will next discuss. In order to perform the discussion we split the cases of odd and even dimension.

Condition (\ref{cond}) in the case of odd dimension is
\begin{eqnarray}
&2^{(n+3)/2}\!-\!n^{2}\!\geqslant\!0
\end{eqnarray}
and its discussion is rather easy: the exponential is always even while the square of an odd number is odd, and therefore we will never be in the optimal case of equality.

The sub-optimal case is given for all dimensions $n\!\geqslant\!11$ while remaining cases are for the $n\!=\!3,5,7,9$ dimensions.

Condition (\ref{cond}) in the case of even dimension is
\begin{eqnarray}
&2^{(n+4)/2}\!-\!n^{2}\!\geqslant\!0
\label{condeven}
\end{eqnarray}
and now an optimal case is possible: it is given for $n\!=\!4,8$ dimensions and by no other dimension as it can be seen from the fact that (\ref{condeven}) can have two solutions at most.

Sub-optimal cases are given for $n\!=\!2$ as well as for all dimensions $n\!\geqslant\!10$ while the remaining case is now given by the single instance of $n\!=\!6$ dimensions solely.

Merging the two instances back together, we have that the degrees of freedom of the spinor field will always be too many to be determined by the independent field equations for $n\!=\!2$ and all $10\!\leqslant\!n$ dimensions in general; within the interval $3\!\leqslant\!n\!\leqslant\!9$ the spinor field is over-determined by too many field equations apart from cases $n\!=\!4,\ 8$ for which degrees of freedom of the spinor field and independent field equations match in a perfect manner.

So the cases $n\!=\!4,\ 8$ are \emph{optimal}. Among the other two cases, the case $n\!=\!2$ and all $10\!\leqslant\!n$ are \emph{sub-optimal} since they are not optimal but there always is the possibility to restrict the spinor field enough to be well determined.

The remaining cases have to be regarded as problematic because there is nothing we can do in order to restore the match between degrees of freedom of the spinor field and the corresponding independent field equations, apart from requiring some arbitrary restrictions on quantities that should otherwise be freely determined, and therefore we will call these cases \emph{quasi-critical} in the following.

We will now proceed to exemplify some of these cases by considering the simplest instance in each class.
\subsection{An optimal case: 4-dimensions}
Among the two optimal cases, we are going to investigate the simplest, that is the case of $4$ dimensions.

This case is the most interesting because it represents what to all empirical evidence appears to be the natural universe; additionally, it is a good starting point because it has already been investigated quite intensively \cite{Fabbri:2016msm}.

In this case, the Clifford matrices $\boldsymbol{\gamma}^{a}$ can be taken for instance in chiral form according to
\begin{eqnarray}
&\boldsymbol{\gamma}^{0}\!=\!\left(\begin{array}{cc}
\!0 & \mathbb{I}\!\\
\!\mathbb{I} &  0\!
\end{array}\right)\ \ \ \ 
\boldsymbol{\gamma}^{A}\!=\!\left(\begin{array}{cc}
\!0 \!&\! \boldsymbol{\sigma}^{A}\!\\
\!-\boldsymbol{\sigma}^{A}\!&\! 0\!
\end{array}\right)
\end{eqnarray}
for $A\!=\!1,2,3$ and with $\boldsymbol{\sigma}^{A}$ Pauli matrices; our convention is to take $\left[\boldsymbol{\gamma}_{a},\!\boldsymbol{\gamma}_{b}\right]\!=\!4\boldsymbol{\sigma}_{ab}$  so that we have
\begin{eqnarray}
&\!\!\!\!\boldsymbol{\sigma}^{0I}\!=\!-\frac{1}{2}\left(\begin{array}{cc}
\!\boldsymbol{\sigma}^{I}\!\!&\!\! 0\!\!\\
\!\!0 &\! -\boldsymbol{\sigma}^{I}\!\!
\end{array}\right)\ \ \ \ 
\boldsymbol{\sigma}^{JK}\!=\!-\frac{i}{2}\varepsilon^{IJK}\left(\begin{array}{cc}
\!\boldsymbol{\sigma}^{I} \!\!&\!\! 0\!\!\\
\!\!0 & \boldsymbol{\sigma}^{I}\!
\end{array}\right)
\end{eqnarray}
for $I,J,K\!=\!1,2,3$ as above. It is also essential to define through $2i\boldsymbol{\sigma}_{ab}\!=\!\varepsilon_{abcd}\boldsymbol{\pi}\boldsymbol{\sigma}^{cd}$ the matrix $\boldsymbol{\pi}$ (which is merely the fifth gamma matrix in a notation without the useless index five, since we are treating the $4$-dimensional case).

It is also important to recall the relationship given by
\begin{eqnarray}
&\boldsymbol{\gamma}_{i}\boldsymbol{\gamma}_{j}\boldsymbol{\gamma}_{k}
\!=\!\boldsymbol{\gamma}_{i}\eta_{jk}\!-\!\boldsymbol{\gamma}_{j}\eta_{ik}
\!+\!\boldsymbol{\gamma}_{k}\eta_{ij}
\!+\!i\varepsilon_{ijkq}\boldsymbol{\pi}\boldsymbol{\gamma}^{q}
\label{id4}
\end{eqnarray}
which is a general spinorial identity.

With the $\boldsymbol{\sigma}_{ab}$ matrices it is possible to construct the complex Lorentz transformation, and the spinorial fields are defined as what transforms according to such a complex Lorentz transformation. With $\boldsymbol{\gamma}_{0}$ we can define the operation $\overline{\psi}\!=\! \psi^{\dagger}\boldsymbol{\gamma}_{0}$ as the spinorial field conjugation, in terms of which the bilinear quantities
\begin{eqnarray}
&M_{ab}\!=\!2i\overline{\psi}\boldsymbol{\sigma}_{ab}\psi\\
&S^{a}\!=\!\overline{\psi}\boldsymbol{\gamma}^{a}\boldsymbol{\pi}\psi\\
&U^{a}\!=\!\overline{\psi}\boldsymbol{\gamma}^{a}\psi\\
&\Theta\!=\!i\overline{\psi}\boldsymbol{\pi}\psi\\
&\Phi\!=\!\overline{\psi}\psi
\end{eqnarray}
are all real tensors as it is easy to demonstrate.

With (\ref{id4}) and these bilinear quantities it is possible to prove the validity of relationships
\begin{eqnarray}
&\frac{1}{2}M_{ab}M^{ab}\!=\!\Phi^{2}\!-\!\Theta^{2}\label{M0}\\
&M_{ik}U^{i}=\Theta S_{k}\label{M1}\\
&M_{ik}S^{i}=\Theta U_{k}\label{M2}
\end{eqnarray}
as well as
\begin{eqnarray}
&U_{a}U^{a}\!=\!-S_{a}S^{a}\!=\!\Theta^{2}\!+\!\Phi^{2}\label{SU21}\\
&U_{a}S^{a}=0\label{SU0}
\end{eqnarray}
called Fierz rearrangements and being spinor identities.

As is discussed in \cite{Fabbri:2016msm}, it is possible to prove that, if not both $\Theta$ and $\Phi$ vanish identically, as it happens in general, then we can always employ some Lorentz transformation to remove redundant components in the spinor field; now because there are up to $6$ available transformations, then there are up to $6$ components that can be removed, and in fact in this case it is possible to remove them all thus reducing the spinor field to $2$ components only: in such a case the spinor field acquires the form
\begin{eqnarray}
&\!\psi\!=\!\phi e^{-\frac{i}{2}\beta\boldsymbol{\pi}}
\boldsymbol{S}\left(\!\begin{tabular}{c}
$1$\\
$0$\\
$1$\\
$0$
\end{tabular}\!\right)
\label{spinor4}
\end{eqnarray}
with $\phi$ and $\beta$ respectively scalar and pseudo-scalar called module and Yvon-Takabayashi angle while $\boldsymbol{S}$ is a generic Lorentz transformation and which is called polar form of the spinorial field. Notice that in the polar form we have
\begin{eqnarray}
&\!\!\!\!M_{ab}\!=\!2\phi^{2}(\cos{\beta}u^{j}s^{k}\varepsilon_{jkab}\!+\!\sin{\beta}u_{[a}s_{b]})
\end{eqnarray}
in terms of
\begin{eqnarray}
&\!\!S^{a}\!=\!2\phi^{2}s^{a}\\
&\!\!U^{a}\!=\!2\phi^{2}u^{a}
\end{eqnarray}
and
\begin{eqnarray}
&\!\Theta\!=\!2\phi^{2}\sin{\beta}\label{b2}\\
&\!\Phi\!=\!2\phi^{2}\cos{\beta}\label{b1}
\end{eqnarray}
as it can be demonstrated very straightforwardly indeed.

The proof can be obtained either by directly performing Lorentz transformations onto the spinor field; or by noticing that, because of identity (\ref{SU21}), it turns out that $U^{a}$ is a time-like vector, so we can always boost its space part away, and then we can always rotate along the third axis the spatial part of $S^{a}$ in general: either way, we get the spinor field in polar form (\ref{spinor4}) in general \cite{Fabbri:2016msm}. Also
\begin{eqnarray}
&u_{a}u^{a}\!=\!-s_{a}s^{a}\!=\!1\\
&u_{a}s^{a}\!=\!0
\end{eqnarray}
showing that $\phi$ and $\beta$ are the only degrees of freedom.

Because spinor fields are complex, in the following we will allow Lorentz transformations $\boldsymbol{S}$ to be accompanied by a generic gauge shift $e^{-i\theta}$ so to write the most general transformation according to $\boldsymbol{S}e^{-i\theta}$ and in this form, we notice that we can always write $(\boldsymbol{S}e^{-i\theta})\partial_{\mu}(\boldsymbol{S}e^{-i\theta})^{-1}$ as
\begin{eqnarray}
&(\boldsymbol{S}e^{-i\theta})\partial_{\mu}(\boldsymbol{S}e^{-i\theta})^{-1}
\!=\!i\partial_{\mu}\theta\mathbb{I}
\!+\!\frac{1}{2}\partial_{\mu}\theta_{ij}\boldsymbol{\sigma}^{ij}
\end{eqnarray}
where $\theta$ is a generic complex phase and $\theta_{ij}\!=\!-\theta_{ji}$ are the six parameters of the Lorentz group; introducing a spin connection $\Omega^{a}_{b\pi}$ as well as a gauge potential $A_{\mu}$ we define
\begin{eqnarray}
&\partial_{\mu}\theta_{ij}\!-\!\Omega_{ij\mu}\!\equiv\!R_{ij\mu}\label{R}\\
&\partial_{\mu}\theta\!-\!qA_{\mu}\!\equiv\!P_{\mu}\label{P}
\end{eqnarray}
which are proven to be real tensors \cite{Fabbri:2016laz}: it is now possible to demonstrate that the Dirac field equations given by
\begin{eqnarray}
&i\boldsymbol{\gamma}^{\mu}\boldsymbol{\nabla}_{\mu}\psi
\!-\!XW_{\mu}\boldsymbol{\gamma}^{\mu}\boldsymbol{\pi}\psi\!-\!m\psi\!=\!0
\end{eqnarray}
with the spinor field in polar form can be written according to the Gordon decomposition in polar form
\begin{eqnarray}
\nonumber
&\!\frac{1}{2}\varepsilon_{\mu\alpha\nu\iota}R^{\alpha\nu\iota}
\!-\!2P^{\iota}u_{[\iota}s_{\mu]}+\\
&+2(\nabla\beta/2\!-\!XW)_{\mu}\!+\!2s_{\mu}m\cos{\beta}\!=\!0\label{dep14}\\
\nonumber
&\!\!R_{\mu a}^{\phantom{\mu a}a}\!-\!2P^{\rho}u^{\nu}s^{\alpha}\varepsilon_{\mu\rho\nu\alpha}+\\
&+2s_{\mu}m\sin{\beta}\!+\!\nabla_{\mu}\ln{\phi^{2}}\!=\!0\label{dep24}
\end{eqnarray}
where $W_{\mu}$ is the torsion axial-vector added for generality and $m$ is the mass of the spinor field. We also have
\begin{eqnarray}
&\nabla_{\mu}s_{i}\!=\!R_{ji\mu}s^{j}\\
&\nabla_{\mu}u_{i}\!=\!R_{ji\mu}u^{j}
\end{eqnarray}
as general constraints on the vector and the axial-vector.

The proof can be obtained multiplying the Dirac field equation by the $\boldsymbol{\gamma}^{a}$ matrices and the conjugate spinorial field and then splitting in real/imaginary parts, as it has been shown in \cite{Fabbri:2016laz}. Notice that (\ref{dep14}, \ref{dep24}) are vector equations, hence specifying all $4$ of the first-order derivatives of the $2$ real degrees of freedom, amounting to a total of $8$ real independent field equations, which are exactly the same number of the original Dirac field equations.

Thus, in $4$-dimensions the spinor field has $2$ real scalar degrees of freedom and there are $8$ real vectorial independent field equations as it is clear when writing everything in terms of the polar form: the $8$ independent field equations specify all $4$ first-order derivatives of the $2$ degrees of freedom that constitute the spinor field precisely.

This is what was expected, as discussed in \cite{Fabbri:2018crr}.
\subsection{A sub-optimal case: 2-dimensions}
Among the many sub-optimal cases, we are again going to investigate the simplest, that is the $2$ dimensions.

The study will proceed exactly as in the case above.

Now, the Clifford matrices $\boldsymbol{\gamma}^{a}$ can be taken to be
\begin{eqnarray}
&\boldsymbol{\gamma}^{0}\!=\!\left(\begin{array}{cc}
\!0 \!&\! 1\!\\
\!1\!&\! 0\!
\end{array}\right)\ \ \ \ 
\boldsymbol{\gamma}^{1}\!=\!\left(\begin{array}{cc}
\!0 \!& -1\!\\
\!1\!&\ \ 0\!
\end{array}\right)
\end{eqnarray}
and because still $\left[\boldsymbol{\gamma}_{a},\!\boldsymbol{\gamma}_{b}\right]\!=\!4\boldsymbol{\sigma}_{ab}$ then
\begin{eqnarray}
&\boldsymbol{\sigma}^{01}\!=\!\frac{1}{2}\left(\begin{array}{cc}
\!1 \!& 0\!\\
\!0\!&\! -1\!
\end{array}\right)
\end{eqnarray}
with $2\boldsymbol{\sigma}_{ab}\!=\!\varepsilon_{ab}\boldsymbol{\pi}$ defining the matrix $\boldsymbol{\pi}$ (the above three matrices are just the three Pauli matrices in the notation without the irrelevant three indices, and again we choose this notation since we are in the $2$-dimensional case).

It is also important to recall the relationship given by
\begin{eqnarray}
&\boldsymbol{\gamma}_{i}\boldsymbol{\gamma}_{j}\boldsymbol{\gamma}_{k}
\!=\!\boldsymbol{\gamma}_{i}\eta_{jk}\!-\!\boldsymbol{\gamma}_{j}\eta_{ik}
\!+\!\boldsymbol{\gamma}_{k}\eta_{ij}\label{id2}
\end{eqnarray}
which is a general spinorial identity.

With the $\boldsymbol{\sigma}_{01}$ matrix it is possible and easy to construct the complex Lorentz transformation, and spinorial fields are defined as what transforms according to such a complex Lorentz transformation. With $\boldsymbol{\gamma}_{0}$ we can define the operation $\overline{\psi}\!=\! \psi^{\dagger}\boldsymbol{\gamma}_{0}$ as the spinorial field conjugation, in terms of which the bilinear quantities
\begin{eqnarray}
&U^{a}\!=\!\overline{\psi}\boldsymbol{\gamma}^{a}\psi\\
&\Theta\!=\!i\overline{\psi}\boldsymbol{\pi}\psi\\
&\Phi\!=\!\overline{\psi}\psi
\end{eqnarray}
are all real tensors as it is easy to demonstrate.

With (\ref{id2}) and these bilinear quantities it is possible to prove the validity of the relationship
\begin{eqnarray}
&U_{a}U^{a}\!=\!\Phi^{2}\!+\!\Theta^{2}
\end{eqnarray}
called Fierz rearrangement and being a spinor identity.

It is easy to see that we can always employ the Lorentz transformation to remove $1$ component in the spinor field, which remains with $3$ components: the spinor field is
\begin{eqnarray}
&\!\psi\!=\!\phi\ e^{i\alpha}e^{-\frac{i}{2}\beta\boldsymbol{\pi}}
\boldsymbol{S}\left(\!\begin{tabular}{c}
$1$\\
$1$
\end{tabular}\!\right)
\label{spinor2}
\end{eqnarray}
with $\phi$, $\alpha$ and $\beta$ respectively being a scalar, a scalar and a pseudo-scalar while $\boldsymbol{S}$ being the Lorentz transformation and which is the polar form. Notice that in polar form
\begin{eqnarray}
&\!\!U^{a}\!=\!2\phi^{2}u^{a}
\end{eqnarray}
and
\begin{eqnarray}
&\!\Theta\!=\!2\phi^{2}\sin{\beta}\\
&\!\Phi\!=\!2\phi^{2}\cos{\beta}
\end{eqnarray}
and where we can witness that the $\alpha$ scalar never appears.

The proof can be obtained by directly performing the Lorentz transformation on the spinor field. We also have
\begin{eqnarray}
&u_{a}u^{a}\!=\!1
\end{eqnarray}
with $\phi$, $\alpha$ and $\beta$ being the only degrees of freedom.

As before, we will allow the Lorentz transformation to be accompanied by the gauge shift, so that we can define the $\theta$ and $\theta_{01}\!=\!-\theta_{10}$ parameters; with $\Omega^{a}_{b\pi}$ and $A_{\mu}$ it can be possible to define $R_{ij\mu}$ and $P_{\mu}$ as above: and now, as before, we can demonstrate that the Dirac field equations with a spinor field in polar form can be written as
\begin{eqnarray}
&-2(P\!-\!\nabla \alpha)^{a}\varepsilon_{a\mu}
\!+\!\nabla_{\mu}\beta\!+\!2mu^{a}\varepsilon_{a\mu}\cos{\beta}=0\label{dep12}\\
&R_{\mu a}^{\phantom{\mu a}a}\!+\!2mu^{a}\varepsilon_{a\mu}\sin{\beta}
\!+\!\nabla_{\mu}\ln{\phi^{2}}\!=\!0\label{dep22}
\end{eqnarray}
with no torsion in two dimensions and in which $m$ is the mass of the spinor field. We also have
\begin{eqnarray}
&\nabla_{\mu}u_{i}\!=\!R_{ji\mu}u^{j}
\end{eqnarray}
as general constraints on the vector alone.

The proof is obtained as above. And as above, we get vector equations, specifying the $2$ first-order derivatives of $\phi$ and $\beta$ as $2$ real degrees of freedom, amounting to a total of $4$ real independent field equations, which are the exact same number of the original Dirac field equations as it is possible to check; nevertheless, there is no equation specifying the first-order derivatives of $\alpha$ which thus is a degree of freedom with no independent field equation.

Thus, in $2$-dimensions the spinor field has $3$ real scalar degrees of freedom and there are $4$ real vectorial independent field equations as it is clear in polar form: the $4$ independent field equations specify the $2$ first-order derivatives of $2$ degrees of freedom but $1$ degree of freedom is left undetermined. Hence, this situation is sub-optimal.

However, it can be made optimal by having the spinor field restricted enough to lose the undetermined degree of freedom: and since such degree of freedom is represented by a phase, we could arrange to have the gauge shift be re-scaled as $\theta\!\rightarrow\!\theta+\alpha$ and the extra phase absorbed away.

Absorbing the phase away is equivalent to require that the gauge charge be equal to zero, which means that what we are considering are spinors of Majorana type solely.

In dimension $2$ Majorana spinors are also Weyl spinors.

As a consequence, in such a case there are more degrees of freedom than independent field equations although the extra degree of freedom could be appropriately removed then rendering the actual number of degrees of freedom exactly the same as the total number of independent field equations, which means that this case is sub-optimal but it can be rendered optimal by having the spinorial fields restricted to be the Majorana-Weyl spinorial fields.

Nevertheless we have to specify that the status of optimal case is recovered only because there was one single extra component that could be reabsorbed with the phase shift and so this procedure works only in $2$ dimensions.

In other dimensions one would have to rely on a different procedure for removing the extra components.
\subsection{A quasi-critical case: 3-dimensions}
Among the five quasi-critical cases, we are again going to investigate the simplest, and that is the $3$ dimensions.

The study will proceed similarly to the cases above.

Now, Clifford matrices $\boldsymbol{\gamma}^{a}$ are taken to be given by
\begin{eqnarray}
&\boldsymbol{\gamma}^{0}\!=\!\left(\begin{array}{cc}
\!0 \!&\! 1\!\\
\!1\!&\! 0\!
\end{array}\right)\ \ \ 
\boldsymbol{\gamma}^{1}\!=\!\left(\begin{array}{cc}
\!0 \!& -1\!\\
\!1\!&\ \ 0\!
\end{array}\right)\ \ \ 
\boldsymbol{\gamma}^{2}\!=\!\left(\begin{array}{cc}
\!i \!& 0\!\\
\!0\!&\! -i\!
\end{array}\right)
\end{eqnarray}
and because $\left[\boldsymbol{\gamma}_{a},\!\boldsymbol{\gamma}_{b}\right]\!=\!4\boldsymbol{\sigma}_{ab}$ then
\begin{eqnarray}
&2\boldsymbol{\sigma}^{ab}\!=\!i\varepsilon^{abc}\boldsymbol{\gamma}_{c}
\end{eqnarray}
with no possibility to define further matrices (once again, the above three matrices are just the three Pauli matrices in the notation with the three indices renamed to match the structure of space-time in the $3$-dimensional case).

It is also important to recall the relationship given by
\begin{eqnarray}
&\boldsymbol{\gamma}_{i}\boldsymbol{\gamma}_{j}\boldsymbol{\gamma}_{k}
\!=\!\boldsymbol{\gamma}_{i}\eta_{jk}\!-\!\boldsymbol{\gamma}_{j}\eta_{ik}
\!+\!\boldsymbol{\gamma}_{k}\eta_{ij}\!+\!i\varepsilon_{ijk}\mathbb{I}\label{id3}
\end{eqnarray}
which is a general spinorial identity as usual.

And as usual, with the $\boldsymbol{\sigma}_{ab}$ matrices we can construct the complex Lorentz transformation, and spinorial fields are defined as what transforms according to such a complex Lorentz transformation. With $\boldsymbol{\gamma}_{0}$ we can define the operation $\overline{\psi}\!=\! \psi^{\dagger}\boldsymbol{\gamma}_{0}$ as the spinorial field conjugation, in terms of which the bilinear quantities
\begin{eqnarray}
&U^{a}\!=\!\overline{\psi}\boldsymbol{\gamma}^{a}\psi\\
&\Phi\!=\!\overline{\psi}\psi
\end{eqnarray}
are all real tensors as it is easy to demonstrate.

With (\ref{id3}) and these bilinear quantities it is possible to prove the validity of the relationship
\begin{eqnarray}
&U_{a}U^{a}\!=\!\Phi^{2}
\end{eqnarray}
called Fierz rearrangement and being a spinor identity.

It is easy to see that we can always employ the Lorentz transformations to remove $3$ components in spinor fields, which remain with $1$ component only: spinor fields are
\begin{eqnarray}
&\!\psi\!=\!\phi \boldsymbol{S}\left(\!\begin{tabular}{c}
$1$\\
$1$
\end{tabular}\!\right)
\label{spinor3}
\end{eqnarray}
with $\phi$ scalar and $\boldsymbol{S}$ Lorentz transformation and which is the polar form. Notice that in polar form
\begin{eqnarray}
&\!\!U^{a}\!=\!2\phi^{2}u^{a}
\end{eqnarray}
and
\begin{eqnarray}
&\!\Phi\!=\!2\phi^{2}
\end{eqnarray}
where every possible phase has disappeared completely.

The proof can be obtained by directly performing the Lorentz transformations on spinor fields. We also have
\begin{eqnarray}
&u_{a}u^{a}\!=\!1
\end{eqnarray}
and as it is clear $\phi$ is the only degree of freedom possible.

As before, we will allow the Lorentz transformations to be accompanied by the gauge shift, so that we can define the $\theta$ and $\theta_{ij}\!=\!-\theta_{ji}$ parameters; with $\Omega^{a}_{b\pi}$ and $A_{\mu}$ it can be possible to define $R_{ij\mu}$ and $P_{\mu}$ as above: once again, we can demonstrate that the Dirac field equations with the spinor field in polar form can be written as
\begin{eqnarray}
&\frac{1}{2}R_{ija}\varepsilon^{ija}\!+\!2P^{k}u_{k}
\!-\!2XW\!-\!2m\!=\!0\label{dep13}\\
&R_{\mu a}^{\phantom{\mu a}a}\!+\!2\varepsilon_{\mu ak}P^{a}u^{k}
\!+\!\nabla_{\mu}\ln{\phi^{2}}\!=\!0\label{dep23}
\end{eqnarray}
in which $W$ is the torsion scalar added for generality and $m$ is the mass of the spinor field. We also have
\begin{eqnarray}
&\nabla_{\mu}u_{i}\!=\!R_{ji\mu}u^{j}
\end{eqnarray}
as general constraints on the vector alone.

The proof is obtained as above, again. But now, quite differently from above, we have a vector equation specifying the $3$ first-order derivatives of $\phi$ as the real degree of freedom, therefore $3$ real independent field equations, which are $1$ short of the $4$ original Dirac field equations, as it is clear to see: the remaining field equation (\ref{dep13}) has been converted into a constraining equation for some of the quantities usually involved in the dispersion relations.

So, in $3$-dimensions the spinor field has one real scalar degree of freedom with $4$ real vectorial independent field equations as it is clear in polar form: the $4$ independent field equations specify the $3$ first-order derivatives of the degree of freedom but they are supplemented with $1$ constraining equation among some quantities that cannot be instead freely given. Therefore, this case is quasi-critical.

Clearly, it is always possible that for situations of specific symmetries the constraining equation is verified by chance. However, its validity is not granted in general.

In fact, to verify the constraint it would be required to restrict quantities on which we have no control at all.

And of course, the number of constraints is even larger for any space-time with more than $3$ dimensions.
\section{Comments}
In the above section we found condition (\ref{cond}) as the condition that has to be verified in order for the spinor field and its field equations to be properly defined: when it is not verified it simply means that the spinorial field has a number of degrees of freedom that extinguishes that of the independent field equations but with some remaining field equation that converts into a constraining equation which cannot be verified in general; when it is verified in the strict sense it means that there are no field equations converting into constraining equations and the independent field equations are as many as the degrees of freedom of the spinor field, while when it is verified in the broad sense the independent field equations are not enough to determine all degrees of freedom of the spinor field.

To better convince the reader, we have provided some well-known examples, one for each case: 1. in $4$ dimensions, the Dirac field equations are equivalent to the pair given by (\ref{dep14}, \ref{dep24}), and this is optimal; 2. in $2$ dimensions, the Dirac field equations are equivalent to the couple of equations given by (\ref{dep12}, \ref{dep22}) where the $\alpha$ degree of freedom does not have an independent field equation, henceforth this case is said sub-optimal; 3. in $3$ dimensions, a Dirac set of field equations is equivalent to the two field equations that are given by the equations (\ref{dep13}, \ref{dep23}) although
\begin{eqnarray}
&\frac{1}{4}R_{ija}\varepsilon^{ija}\!+\!P^{k}u_{k}\!-\!XW\!-\!m\!=\!0
\end{eqnarray}
is actually a constraint, and as a consequence this problematic situation has been indicated as quasi-critical.

We had noticed that case 2. has problems which could be solved by restricting the generality of the spinor field, although the method used in the above instance was also not of general applicability: the extra degree of freedom was a single phase and we could use the gauge transformation to gauge the phase away, but this way out clearly works only for a single extra degree of freedom; for sub-optimal cases of larger dimensions this specific restriction would not be enough, and further restrictions will have to be implemented. Just the same, in different dimensions there will be correspondingly different methods to follow in order to restrict the spinor: for example, like in the $2$ dimensional case spinors could be of Majorana type and of Weyl type simultaneously, also in $10$ dimensions such Majorana-Weyl spinor can be defined, and therefore like in the $2$ dimensional case, also in $10$ dimensions it may be easier to recover the optimal condition. In dimensions in which this is not possible, we can only require spinors to be either Majorana spinors or Weyl spinors, and each of these conditions taken independently may not be enough to recover the situation of optimal case for the spinor.

Nevertheless, the condition of being a Majorana spinor or a Weyl spinor is not immediately applicable in general in the context of the polar form: at least in the case of $4$ dimensions \cite{Fabbri:2016msm,Fabbri:2016laz,Fabbri:2018crr}, it is a known fact that spinors can be split in two classes, according to whether they are regular (at least one of the two bi-linear scalars different from zero) or singular (both bi-linear scalars vanishing), with general Dirac spinors belonging to the former class and Majorana and Weyl spinors belonging to the latter class. Both classes admit a polar decomposition, but the two polar decompositions are not equivalent, so that they have to be studied independently, and therefore we can not apply to one the conditions obtained for the other.

Because in this paper we have focused on the polar decomposition of the regular spinors, seeing what happens when implementing the condition of singularity of spinors would require an extension of the analysis which would exceed the reasonable limits of the paper.

As for the case 3. we regard it to be worse, because in this case we would have to release constraints on quantities that cannot be fixed a priori: symmetries, or other hypotheses, may come to rescue. It is not impossible that 
\begin{eqnarray}
&\frac{1}{4}R_{ija}\varepsilon^{ija}\!+\!P^{k}u_{k}\!-\!XW\!-\!m\!=\!0\label{constr}
\end{eqnarray}
be verified by some tailored assumption.

But quite clearly, solvability is precluded, at least if we care to consider the most general of circumstances.

As it stands, we will retain the optimal case as the only meaningful one, so far as generality is concerned.
\section{Conclusion}
In this paper, we have taken a great advantage of the polar form of spinor fields and their field equations as a methodological technique for the analysis of the match between degrees of freedom and independent field equations in various dimensions: we have witnessed that the degrees of freedom of the spinor field are too many to be determined by the independent field equations in dimensions $n\!=\!2$ and all $10\!\leqslant\!n$ dimensions; we have seen that the degrees of freedom of the spinor field are too few for the independent field equations which therefore develop constraints in $n\!=\!3,5,6,7,9$ dimensions. However, it was also noticed that the degrees of freedom of spinor fields are determined by the independent field equations with no additional constraint in $n\!=\!4,8$ dimensions. We have discussed how developing constraints brings general lack of solvability, a situation we called quasi-critical case; and that the circumstance where some degrees of freedom are not determined is also problematic, giving a situation we called sub-optimal. And we have concluded that the exact match is the only instance in which no problems arise in general, which we called optimal case. We showed that in sub-optimal cases constraints can be implemented to recover optimal case although the procedure is in general dimensional-dependent, and we have discussed that some constraints like (\ref{constr}) may happen to be satisfied although dealing with quasi-critical cases is a much more arbitrary task to be achieved in general. Of course, optimal cases do not require any adjustment. This is why they can be treated with no issue even in their most general instance.

It is important to stress that we are not saying that a quasi-critical as well as a sub-optimal case should occupy no place in any study; but such non-optimal cases do have issues that have to be treated. A possible investigation is by having the algebraic study extended as to include the differential study, since in this case the existence of any non-physical component is translated into an unbalance of the field equations, and the clearest way to see this is in polar form, as in this way all unbalanced field equations are written explicitly. The polar form of the Dirac spinor field equations is useful also in optimal cases as a way to write field equations in terms of real tensors alone.

As final and intriguing remark, we notice that the Bott periodicity, characterizing spinor representations given in terms of Clifford algebras, seems broken at a differential level, when considering the Dirac spinorial field equations and as is clearest when employing the polar form.

Spinor fields represent one wide and deep sector of research that in the literature has been extensively and intensively investigated from the point of view of Clifford algebras but which reveals notable aspects when studied with the Dirac field equations in polar form.
\begin{acknowledgments}
I would like to thank Professor Michael E. Peskin for the interesting discussions we had at the GGI 2018.
\end{acknowledgments}

\end{document}